\documentclass{elsarticle}
\usepackage{graphicx}
\biboptions{sort&compress}

\def\fm{\mathrm{fm}}
\def\mev{\mathrm{MeV}}
\def\gev{\mathrm{GeV}}

\def\ri{\mathrm{RI}}

\def\lqcd{\Lambda_\mathrm{QCD}}
\def\msbar{\overline{\mathrm{MS}}}
\def\lsim{\raise0.3ex\hbox{$<$\kern-0.75em\raise-1.1ex\hbox{$\sim$}}}
\def\gsim{\raise0.3ex\hbox{$>$\kern-0.75em\raise-1.1ex\hbox{$\sim$}}}
\def\fm{\mathrm{fm}}

\newcommand{\dbs}{}
\newcommand{\GeV}{\,\mathrm{GeV}}
\newcommand{\MSbar}{\overline{\mathrm{MS}}}

\begin{document}

\begin{frontmatter}

\title{Lattice QCD at the physical point: light quark masses\\[3.3cm]
\large{Budapest-Marseille-Wuppertal Collaboration}\\[-3.8cm]}

\author[w,j]{S.~Durr}
\author[w,j,b]{Z.~Fodor}
\author[w,m]{C.~Hoelbling}
\author[b,w]{S.D.~Katz}
\author[w,j]{S.~Krieg}
\author[w]{T.~Kurth}
\author[m]{L.~Lellouch}
\author[j,w]{T.~Lippert}
\author[w]{K.K.~Szabo}
\author[m]{G.~Vulvert}

\address[w]{Bergische Universit\"at Wuppertal, Gaussstr.\,20, D-42119 Wuppertal, Germany.}
\address[m]{Centre de Physique Th\'eorique\footnote{CPT is research unit UMR 6207 of the CNRS and of the universities Aix-Marseille I, Aix-Marseille II and Sud Toulon-Var, and is affiliated with the FRUMAM.},
Case 907, Campus de Luminy, F-13288 Marseille, France.}
\address[j]{J\"ulich Supercomputing Centre, Forschungszentrum J\"ulich,
D-52425 J\"ulich, Germany.}
\address[b]{Institute for Theoretical Physics, E\"otv\"os University,
H-1117 Budapest, Hungary.}

\date{\today}

\begin{abstract}
Ordinary matter is described by six fundamental parameters: three couplings (gravitational, electromagnetic and strong) and three masses: the electron's ($m_e$) and those of the up ($m_u$) and down ($m_d$) quarks.  An additional mass enters through quantum fluctuations: the strange quark mass ($m_s$).  The three couplings and $m_e$ are known with an accuracy of better than a few per mil. Despite their importance, $m_u$, $m_d$ (their average $m_{ud}$) and $m_s$ are far less accurately known. Here we determine them with a precision below 2\% by performing ab initio lattice quantum chromodynamics (QCD) calculations, in which all systematics are controlled.  We use pion and quark masses down to (and even below) their physical values, lattice sizes of up to 6~fm, and five lattice spacings to extrapolate to continuum spacetime.  All necessary renormalizations are performed nonperturbatively.
\end{abstract}

\end{frontmatter}

The masses of the up, down and strange quarks cannot be measured using standard experimental methods.  The strong interaction confines quarks within hadrons (e.g.\ protons) in such a way that a single quark cannot be isolated. Moreover, the strength of the interaction is such that the mass of a hadron is not the simple sum of the masses of the quarks it contains. Rather it is provided by complicated nonperturbative dynamics (e.g. \cite{Durr:2008zz}). This confinement mechanism is the low energy counterpart of the strong interaction's asymptotic freedom~\cite{Gross:1973id,Politzer:1973fx}, by which the interactions between quarks and gluons weaken as their relative momenta are increased.

Interestingly enough, the experimental data for $m_u$, $m_d$ and $m_s$ has been available for about sixty years (the pion and kaon were discovered in the late 1940's and the proton already 30 years before).  Even the theory of the strong interaction, QCD, which--in principle--completely describes bound states of light quarks, has been known for almost four decades \cite{Fritzsch:1973pi}. The fact that such a fundamental question has remained poorly answered despite the available experimental and theoretical knowledge is related to the computational difficulties one encounters when trying to solve the underlying equations in the domain of interest. The only known systematic technique to solve them is lattice QCD \cite{Wilson:1974sk,Creutz:1980zw}. Several decades of theoretical, algorithmic and hardware development have been necessary to reach the level at which the light quark masses can be determined reliably. This determination is the goal of the present paper.

For many years calculations were done in the quenched approximation. Although this approach omits the most computationally demanding part of a full QCD calculation --the quark determinant obtained after integrating over the fermion fields-- a controlled determination of the strange quark mass in this approximation (with $m_u$=$m_d$=$m_s$ equal to about half the physical $m_s$) took about 20 years~\cite{Garden:1999fg}. Moreover, the physics of the $u$ and $d$ quarks remained inaccessible, because the quenched approximation, an uncontrolled truncation of QCD, distorts the small quark mass behavior~\cite{Bernard:1992mk,Sharpe:1992ft}.

A very important step in the determination of light quark masses was made with the inclusion of $u$ and $d$ sea quark effects ($N_f{=}2$) \cite{Eicker:1997ws,AliKhan:2001tx,Aoki:2002uc,Gockeler:2004rp,DellaMorte:2005kg}. But even there, physical $m_{ud}$ remained elusive, this time for algorithmic reasons. A first breakthrough was made by the MILC collaboration \cite{Aubin:2004ck}, which used an $N_f{=}2{+}1$ staggered fermion formulation to include strange sea quark effects, pushing calculations to smaller light quark masses, finer lattices and larger volumes. Updates from calculations with sea pion masses down to $258\,\mathrm{MeV}$ (given by the RMS average over taste partners for $a{=}0.06$~fm--their lightest valence pion is 177~MeV at $a{=}0.09$~fm)~\cite{Bazavov:2009tw} and on even finer lattices are presented in \cite{Bazavov:2009bb,Bazavov:2009tw}.  On a subset of the MILC configurations, the HPQCD collaboration has obtained indirectly $m_s$ and $m_{ud}$ via the $m_c/m_s$ ratio \cite{Davies:2009ih,McNeile:2010ji}. Due to their use of quenched and partially quenched charmed and strange quarks with a non-unitary staggered formalism and to their error estimates on the input quantities that they use ($m_c$ and $r_1$), this work does not fulfill our conditions for a controlled ab initio calculation (see below). Recently, also ETMC ($N_f{=}2$) \cite{Blossier:2010cr} and RBC-UKQCD ($N_f{=}2{+}1$) \cite{Aoki:2010dy} have presented results with $M_\pi{\gsim}270\,\mev$ and significantly larger error bars.  All $N_f$=2+1 results for $m_s$ and $m_{ud}$ (except for those of the very recent \cite{Aoki:2010dy}) were combined into world averages in \cite{Colangelo:2010et}, which also reviews $N_f$=2 and non-lattice results. Our results are in complete agreement with these averages, albeit with uncertainties smaller by more than a factor of $5$.

A second breakthrough came recently when it was shown that improvements in algorithms \cite{Hasenbusch:2001ne,Luscher:2005rx} allowed the use of Wilson and domain wall fermions, which are free from the open questions of the rooted staggered approach, for ab initio calculations (e.g. \cite{Durr:2008zz,Allton:2008pn}) and even for reaching physically light $m_{ud}$, albeit in small volumes and at a single lattice spacing~\cite{Aoki:2008sm}.

All previous lattice calculations of $m_{ud}$ and $m_s$ have neglected one or more of the six ingredients which we believe are most important for a full and controlled calculation:

{\it 1. The inclusion of the up (u), down (d) and strange (s) quarks in the fermion determinant with an exact algorithm (i.e. with no integration error) and with an action whose universality class is QCD.}  Rooted staggered fermions provide a numerically efficient way to investigate nonperturbative QCD. However, this discretization is neither local nor unitary for $a{>}â0$, making it difficult to show that it leads to QCD in the continuum limit (please see \cite{Bazavov:2009bb} for another point of view).  Here we use, instead, $N_f$=2+1 Wilson fermions with local improvement terms which do not affect the continuum limit.
 
{\it 2. Controlled interpolations and extrapolations of the results to physical quark masses.} Practically it means reaching pion masses as small as 200~MeV (clearly the value depends on the problem and on the required accuracy) or most preferably simulating at the physical mass point itself.  At three of our lattice spacings we use physical (or even smaller) light quark masses.

{\it 3. Large volumes to guarantee small finite-size effects.} Our finite volume corrections are tiny (we use volumes up to 6~fm).  Nevertheless we correct for them.

{\it 4. Controlled extrapolations to the continuum limit.} This requires that calculations be performed at no less than three values of the lattice spacing, to check whether the scaling region is reached. We use five lattice spacings between 0.116 and 0.054~fm, thereby gaining full control on the continuum extrapolation.

{\it 5. Nonperturbative treatment in all steps.} We obtain our primary results ($m_{ud}$ and $m_s$ in the RI scheme at 4~GeV) in a completely nonperturbative manner. In particular, we eliminate all truncation errors associated with the often used perturbative renormalization.

{\it 6. Input parameters.}  The parameters of the theory (scale and quark masses) should be fixed with well measured observables whose error bars are undisputed and whose connection to experiment is transparent and contains no hidden assumptions. To that end we use $M_\pi$, $M_K$ and $M_\Omega$ exclusively.  The influence of their error bars is negligible on our final uncertainties. Taking instead derived quantities, like $m_c$ and $r_1$ as is done in \cite{Davies:2009ih,McNeile:2010ji}, while fine in principle, can be problematic in practice. The error assigned to the input quantity $m_c$ in \cite{McNeile:2010ji} is smaller by a factor 13 than that of the necessarily conservative Particle Data Group value \cite{Nakamura:2010zzi}.  Similarly, due to the difficulties in estimating its systematic uncertainty, the continuum value of $r_1$ (and the related $r_0$) obtained by different groups shows significant differences (e.g. 2.3 combined sigma between \cite{Bazavov:2009bb} and \cite{Aoki:2010dy}).

In this paper we determine $m_{ud}$ and $m_s$, while fulfilling all of the above conditions.  This determination requires two, apparently straightforward, calculations. First we compute hadron masses for tuning the quark masses to their physical values.  Then we determine the renormalization constant to convert the bare quark masses to finite quantities in the continuum limit.

\begin{figure}[t]
\centerline{\includegraphics*[width=8.0cm]{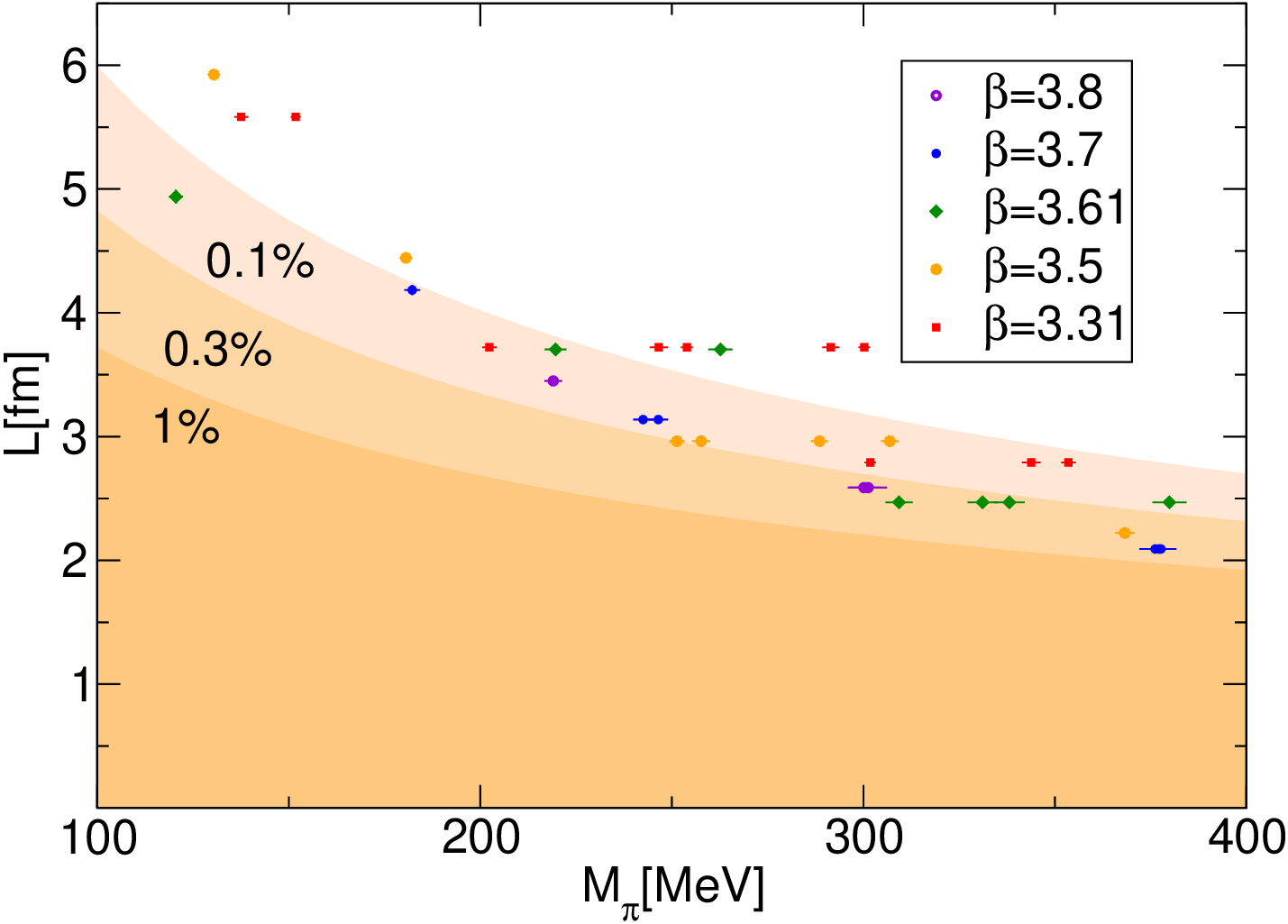}}
\caption{\dbs
\label{landscape} 
Summary of our simulation points. The pion masses and the spatial sizes of the lattices are shown for our five lattice spacings. The percentage labels indicate regions, in which the expected finite volume effect \cite{Colangelo:2003hf} on $M_\pi$ is larger than 1\%, 0.3\% and 0.1\%, respectively. This effect is smaller than about 0.5\% for all of our runs and, as described, we corrected for it. Error bars are statistical.  }
\end{figure}

We now list the most important steps of our work:

{\it (i) Production of the $N_f$=2+1 gauge field ensembles.} We use a Symanzik improved gauge action and 2-level HEX \cite{Hasenfratz:2001mi,Morningstar:2003gk,Capitani:2006ni} smeared clover fermions, with $m_s$ held close to its physical value. Gauge field configurations for 47 different values of the parameters ($\beta{=}6/g^2$, $am_{ud}$ and $am_s$) were produced (c.f. Fig.~\ref{landscape} for our $M_\pi{<}400\,\mev$ $N_f=2+1$ data).

We used five lattice spacings ($a$$\approx$0.116, 0.093, 0.077, 0.065 and 0.054~fm), which are the basis for the continuum extrapolation. As we will see, the difference between the results obtained on the finest lattice and those in the continuum limit is ${\sim}3\%$, whereas between those of the coarsest lattice and the continuum limit is ${\sim}10\%$.

At two pion mass points we carried out detailed finite $V$ analyses, which give us a full understanding of the finite $V$ corrections, as well as their $M_\pi$ dependence \cite{Durr:2010aw}.  In our calculations, we have $M_\pi L{\gsim}4$ and/or $L{\gsim}5\,\fm$, so that the limit $V{\rightarrow}\infty$ can be taken safely. The difference between the results obtained directly on our large lattices and those in the $V{\rightarrow}\infty$ limit is below the five per mil level. Furthermore, for $M_\pi{<}200$~MeV, which is most relevant for our final result, these corrections are even smaller, namely on the one per mil level (see Fig.~\ref{landscape}).

In our calculations $M_\pi$ ranges from ${\approx}380$ down to ${\approx}120\,\mev$ (for three of the five lattice spacings we tuned $M_\pi$ to the vicinity of $135\,\mev$ and for the two finest lattices, the smallest $M_\pi$ are around 180 and $220\,\mev$, respectively). Bracketing the physical mass point allows us to circumvent potentially troublesome chiral extrapolations. We perform calculations with $m_s$ values slightly below and above the physical mass, allowing a straightforward interpolation.

{\it (ii) Hadron and bare quark mass calculations.}  The pion and kaon masses are used to fix $m_{ud}$ and $m_s$ respectively, with $M_\Omega$ providing the overall scale. We take $M_\pi{\simeq}135\,\mev$, $M_K{\simeq}495\,\mev$ and $M_\Omega{\simeq}1672\,\mev$ as input parameters \cite{Durr:2010aw}. The calculation of hadron masses and the ``mass independent scale setting'' follows that of \cite{Durr:2008zz}. All three hadron masses receive finite volume corrections, falling off exponentially with $M_\pi L$ \cite{Luscher:1985dn}. Even though these corrections are tiny, they are corrected for \cite{Durr:2010aw}. In addition to the hadron masses, the unrenormalized partially conserved axial current (PCAC) quark masses are determined.

{\it (iii) Renormalization of the bare quark masses.} In addition to the PCAC masses discussed above, the bare $m_{ud}$ and $m_s$ in the Lagrangian also provide a measure of the quark masses used in our simulations. Once suitably renormalized, these two definitions yield quark masses which must agree in the continuum limit.

While the PCAC masses renormalize multiplicatively, the bare Lagrangian masses require an additional additive renormalization. In the difference $d{\equiv} m_s^\mathrm{bare}-m_{ud}^\mathrm{bare}$, this additive renormalization is eliminated. Moreover, the multiplicative renormalization factors cancel in the ratio $r{\equiv} m_s^\mathrm{PCAC}/m_{ud}^\mathrm{PCAC}$. To obtain fully renormalized quantities, we must still multiply $d$ by $1/Z_S$, the inverse of the scalar density renormalization factor. From the renormalized mass difference $d/Z_S$ and the renormalization independent ratio $r$ we obtain $m^\mathrm{ren}_{ud}=(d/Z_S)/(r-1)$ and $m^\mathrm{ren}_s=(r d/Z_S)/(r-1)$ in the unimproved case.  Our final analysis is tree-level ${\cal O}(a)$ improved with slightly more complicated formulae (see Sec.~11.2 of \cite{Durr:2010aw}).

To compute $Z_S$ nonperturbatively in the RI scheme, we apply the Rome-Southampton (RS) method \cite{Martinelli:1994ty} with tree-level improvement, augmented with nonperturbative running. Our procedure eliminates the possible difficulties of the RS method on coarser lattices. Since the RI scheme is defined in the $N_f{=}3$ (i.e. with three degenerate quarks) chiral limit, we generate $N_f{=}3$ configurations at our five lattice spacings and, for each $\beta$, at four or more values of $m_q$ to allow an extrapolation to the massless limit. Thus for each $\beta$ we compute the renormalization constant $Z_S^\ri(\beta,\mu)$, at renormalization scale $\mu$, as described in \cite{Durr:2010aw}. The RS procedure defines a valid renormalization scheme as long as $\mu{\ll}\pi/a$. However, only if $\mu{\gg}\lqcd$ can the results be converted perturbatively to other schemes (including intrinsically perturbative schemes such as $\msbar$) or be used in a perturbative context. On coarser lattices, it is difficult to simultaneously satisfy both constraints on $\mu$.  To solve this difficulty we first determine the quark masses at $\mu=1.3$ and $2.1\,\gev$, then apply continuum nonperturbative running, as defined in \cite{Constantinou:2010gr}, up to $\mu'=4\,\gev$. 

{\it (iv) Combined analysis of mass and lattice spacing dependence.} For the masses, two strategies, called ``Taylor fit'' and ``chiral fit'' \cite{Durr:2008zz} are applied. Clearly, the results of these fits are dominated by the results at the physical point. In the analysis, two different pion mass ranges are used, namely $M_\pi{<}340,380\,\mev$.

\begin{figure*}[t]
\centerline{\includegraphics*[width=5.5cm]{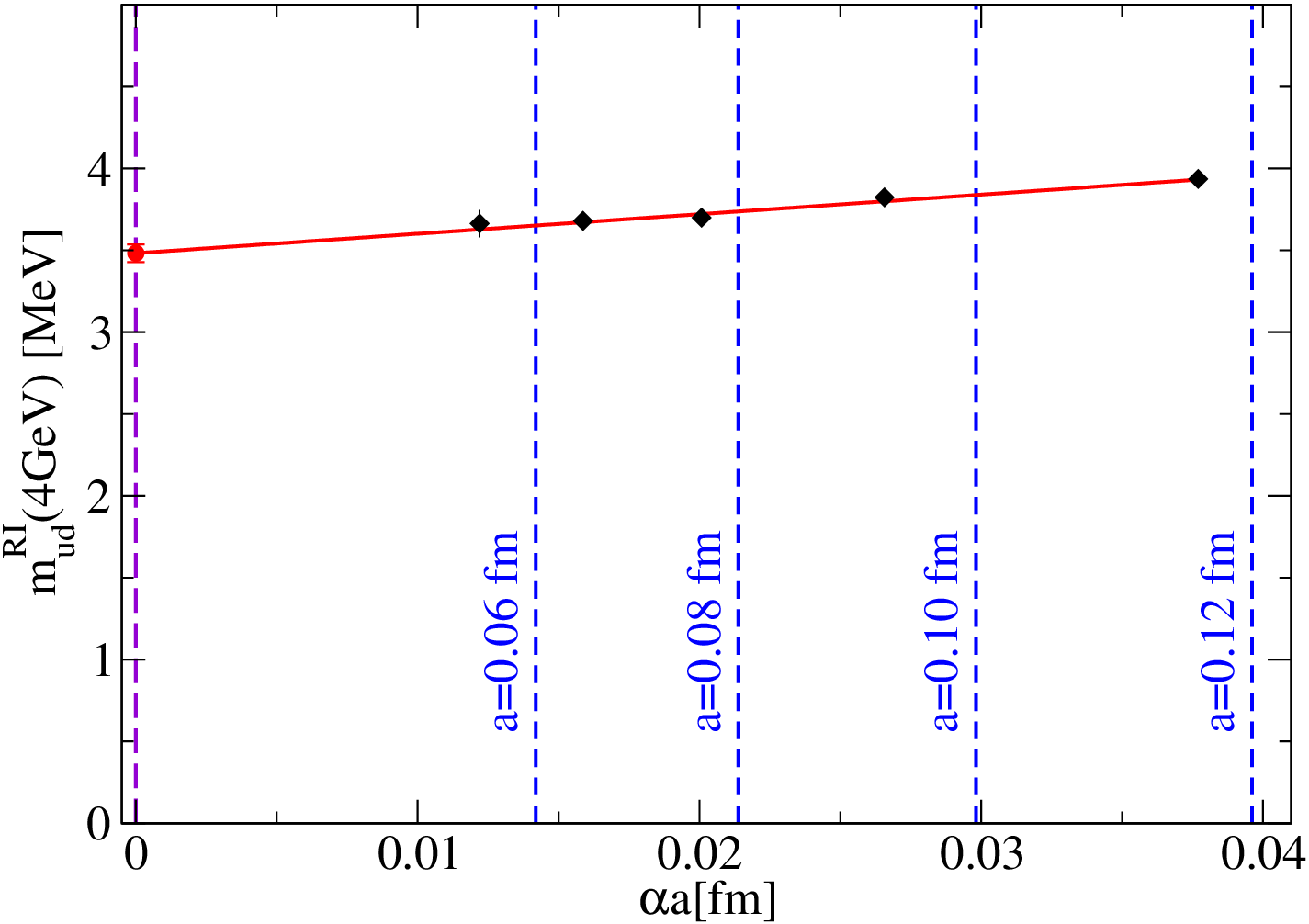}\hspace*{4mm}
\includegraphics*[width=5.7cm]{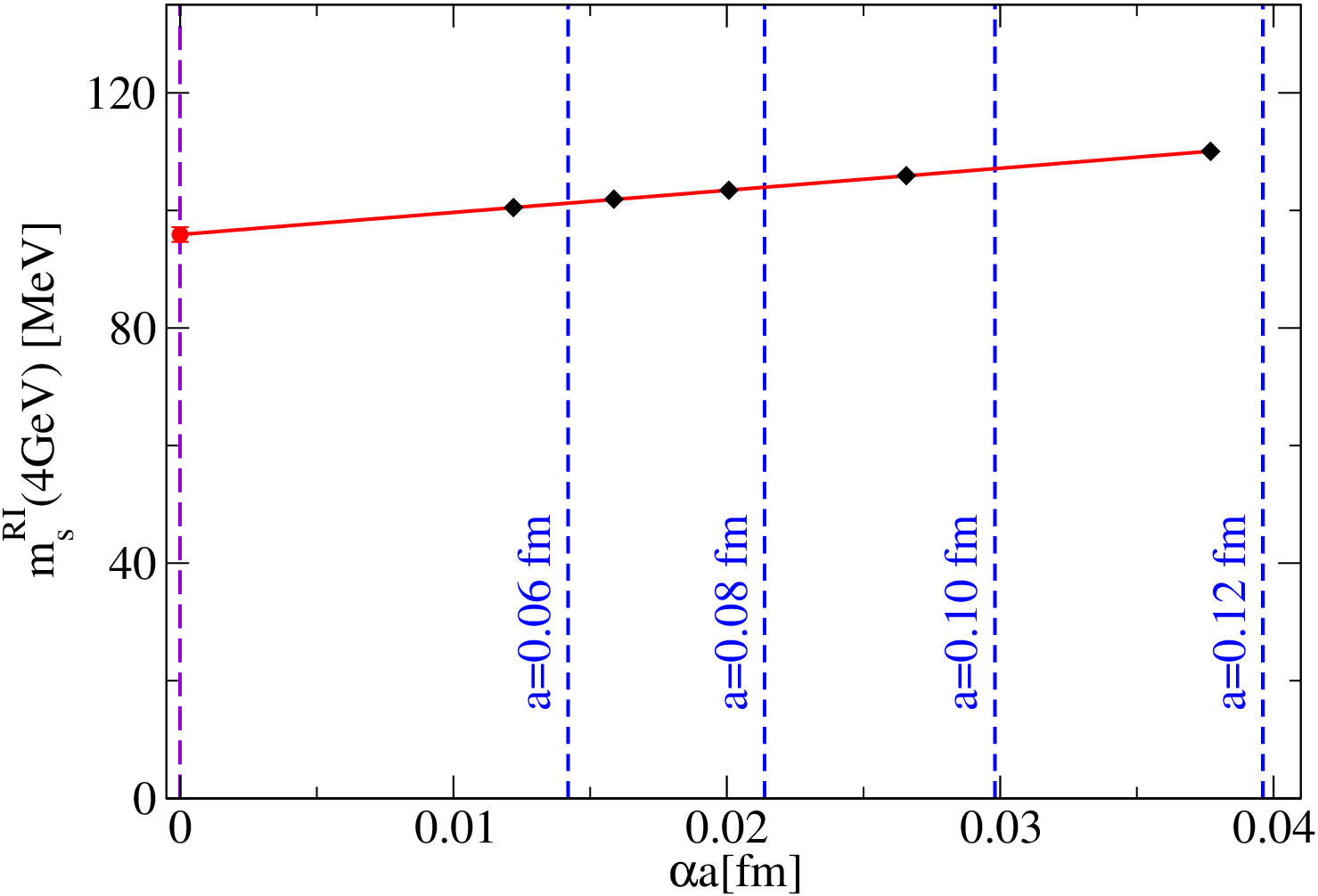}\hspace*{4mm}
\includegraphics*[width=5.5cm]{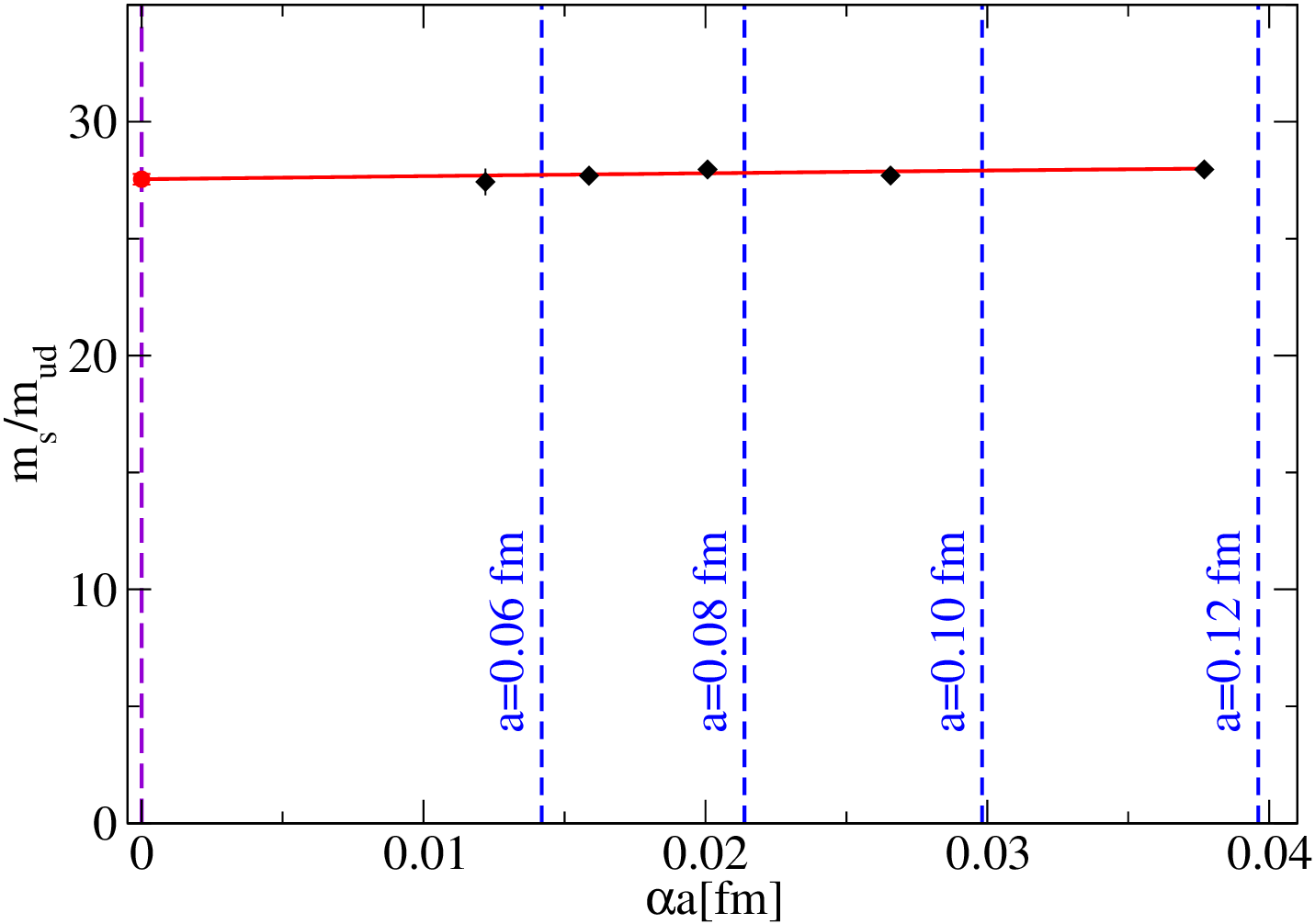}\hspace*{4mm}}
  \caption{\dbs
  \label{mq_final} Continuum extrapolation of the average up/down quark mass,
   of the strange quark mass and of the ratio of the two. The errors of the individual points, which are statistical only here, are smaller than the symbols in most of the cases. The only exceptions are the light quark mass and its ratio to the strange quark mass at the two finest lattice spacings. These exceptions underline the importance of using physical quark masses to reach a high accuracy.  }
\end{figure*}

The strange and average up-down quark masses renormalized in the RI scheme at 4~GeV are extrapolated to the continuum and interpolated to the physical mass point.  In these fits, we include terms to correct linear ($\alpha_s a$) or quadratic ($a^2$) effects. A combined mass and lattice spacing fit is carried out. We show the continuum extrapolation for $m_{ud}$ and $m_s$ in the RI scheme at 4~GeV, as well as their ratio, in Figure \ref{mq_final}.  In order to control the systematic uncertainties we carry out 288 such analyses \cite{Durr:2010aw}.  The figure depicts results from one analysis with one of the best fit qualities.

Our procedure yields the RI quark masses $m_{ud}$ and $m_s$, with statistical and fully controlled systematic errors.  These results do not rely on perturbation theory and from them it is straightforward to obtain the quark masses in other commonly used frameworks such as renormalization group invariant (RGI) and $\msbar$ \cite{Chetyrkin:1999pq} ones. Moreover in \cite{Durr:2010aw}, we show that 4~GeV in the RI scheme is sufficiently large that the perturbative running required to obtain RGI masses and the matching to the $\msbar$ scheme at 2~GeV both yield subdominant uncertainties.

The determination of the individual up and down quark masses at the physical point is in principle possible using exclusively lattice simulations. To that end one may include the electromagnetic U(1) gauge field into the lattice framework, as was done recently in~\cite{Blum:2010ym}. Such a project goes beyond the scope of the present paper, which deals with QCD only. Nevertheless our precise $m_s$ and $m_{ud}$ values can be combined \cite{Durr:2010aw} with model-independent results based on dispersive studies of $\eta{\rightarrow}3\pi$ decays to derive the individual up and down quark masses (c.f. Tab. \ref{mass_results}). In this approach the relationship between the input parameters and experiments is not as transparent as for the determination of $m_s$ and $m_{ud}$ (see condition 6 above).

\begin{table}[t]
\centering
\begin{tabular}{lrrr}
\hline
\hline
& $\mbox{RI}(4\GeV)$ & $\mbox{RGI}$$\qquad$  &$\MSbar(2\GeV)$  \\\hline
$m_s$ & 96.4(1.1)(1.5) & 127.3(1.5)(1.9) & 95.5(1.1)(1.5) \\
$m_{ud}$ &3.503(48)(49) & 4.624(63)(64)  & 3.469(47)(48) \\\hline
$m_u$ & 2.17(04)(10) & 2.86(05)(13) & 2.15(03)(10) \\
$m_d$ & 4.84(07)(12) & 6.39(09)(15) & 4.79(07)(12) \\
\hline
\hline
\end{tabular}
\caption{\dbs
\label{mass_results} 
Renormalized quark masses in the RI scheme at 4~GeV, and after conversion to RGI and the $\MSbar$ scheme at 2~GeV. The RI values are fully nonperturbative, so the first column is our main result. The first two rows emerge directly from our lattice calculation. The last two include additional dispersive information.  }
\end{table}

We have performed a lattice QCD determination of the light quark masses which includes all of the ingredients that we believe are required to achieve full control over systematic errors. In particular, we have eliminated the need for difficult chiral extrapolations by performing simulations all the way down to the physical pion mass (and even below); gained control over the necessary continuum extrapolation by performing simulations at five lattice spacings down to $a=0.054\,\fm$; eliminated perturbative uncertainties by performing the renormalization and running fully nonperturbatively; and controlled the infinite volume extrapolation by working with lattice sizes up to $6\,\fm$.

The precision reached for $m_s$ and $m_{ud}$ is somewhat below the 2\% level and for $m_s/m_{ud}{=}27.53(20)(08)$, which is scheme-independent, it is better than 1\%. For $m_u$ and $m_d$, which include additional dispersive information, it is about 5\% and 3\%, respectively. Despite their use of significantly different methods, MILC~\cite{Bazavov:2009tw}, RBC~\cite{Aoki:2010dy} and HPQCD \cite{McNeile:2010ji}, the three collaborations which have performed the most extensive $N_f{=}2{+}1$ computations besides ours, obtain results for these masses which are at most 1.5 combined standard deviations away from ours.

Our results provide precise and reliable input for phenomenological calculations which require light quark mass values. They highlight the progress that has been achieved since the early days of quark mass determinations \cite{Gasser:1982ap} by showing that phenomenologically relevant lattice QCD calculations can now be carried out bracketing the physical values of the light quark masses.

The details of this work can be found in \cite{Durr:2010aw}.

{\bf Acknowledgments} We used HPC resources from FZ J\"ulich and from GENCI-[IDRIS/CCRT] grant 52275, as well as clusters at Wuppertal and CPT.  This work is supported in part by EU grants I3HP, FP7/2007-2013/ERC n\textsuperscript{\tiny 0} 208740, MRTN-CT-2006-035482 (FLAVIAnet), DFG grants FO 502/2, SFB-TR 55 and CNRS grants GDR 2921 and PICS 4707.


\end{document}